\documentclass{article}
\usepackage[usenames,dvipsnames]{color}
\usepackage{hyperref}
\usepackage{slashed}
\usepackage{amssymb}
\usepackage{latexsym}
\usepackage{amsmath}
\usepackage{authblk}

\newcommand{\fb}{\ensuremath{\bar{f}}}
\newcommand{\Hb}{\ensuremath{\bar{H}}}
\newcommand{\Rb}{\ensuremath{\bar{R}}}
\newcommand{\abar}{\ensuremath{\bar{a}}}

\newcommand{\grad}{\ensuremath{ \vec{\nabla} }}
\newcommand{\PhiGI}{\ensuremath{\hat{\Phi}}}
\newcommand{\PsiGI}{\ensuremath{\hat{\Psi}}}
\newcommand{\FRW}{\ensuremath{{\rm FRW}}}
\newcommand{\dS}{\ensuremath{{\rm dS}}}
\newcommand{\SSS}{\ensuremath{{\rm SSS}}}

\begin{document}

\title{Conditions for equivalence of static spherically symmetric spacetimes to almost Friedman-Robertson-Walker}

\author[1,2]{Constantinos Skordis}
\affil[1]{ Institute of Physics, Czech Academy of Sciences, Prague, Czech Republic.~\footnote{skordis@fzu.cz}}
\affil[2]{ Department of Physics, University of Cyprus, 1, Panepistimiou Street, 2109, Aglantzia, Cyprus.}

\maketitle

\abstract
We determine the conditions under which a static spherically symmetric spacetime metric is equivalent to a perturbed Friedman-Robertson-Walker metric.
We construct the correspondence between the two metrics and discuss a simple application.

\section{Introduction}
Exact solutions to the field equations of gravitational theories, although often having a limited realm of validity,
 are important in that they provide valuable intuition on the behaviour of the theory in idealized situations.
 Exact solutions also serve as starting points for the study of small fluctuations away from such idealized situations.
Two standard examples, relevant to our work, are the static spherically symmetric (SSS) spacetimes and the Friedman-Robertson-Walker (FRW) spacetimes.
The former has led to the study of quasi-normal modes of black holes as perturbations 
 of the Schwarzschild metric while the latter has advanced our understanding of cosmology.

In General Relativity (GR) and to a lesser extend in other theories of gravity~\cite{CliftonFerreiraPadillaSkordis2011},
 numerous exact solutions have been found~\cite{StephaniEtAl2003,GriffithsPodolsky2009}. Due to general covariance
 --a fundamental property of all geometric gravitational theories-- the same solution may be written in different coordinate 
systems. As such, it is not immediately obvious whether two given metrics represent the same solution
and a number of  methods have been developed for deciding whether that is the case, or not, in a coordinate independent manner
 (see, for instance, chapter 9 of ~\cite{StephaniEtAl2003}). However, these methods are in general laborious, involving for instance 
 multiple derivatives of the Riemann tensor. If the metrics are sufficiently simple and if one has reasons to believe that they represent the same solution, 
it may be sometimes easier to construct a coordinate transformation between them.  If such a coordinate transformation can be found 
 it would establish the equivalence of the solutions.

A particularly interesting solution of GR in the presence of a spatially homogeneous (but time dependent) density 
and a spatially inhomogeneous pressure was given by McVittie~\cite{McVittie1933}.
The physical interpretation of this solution has been a subject of debate (see ~\cite{Nolan1998,Nolan1999a,Nolan1999b,KaloperKlebanMartin2010} 
and references therein). In the case that the Hubble parameter $H(t)$ derived from the metric asymptotes to a constant,
~\cite{KaloperKlebanMartin2010} show that the solution describes a black hole embedded in an expanding universe. Interestingly, 
when $H(t)$ is exactly constant the solution becomes the Schwarzschild-de Sitter (SdS) solution, that is, 
 the exterior solution of an isolated non-rotating spherical mass $M$ embedded in a universe with a cosmological constant $\Lambda$. 
That is, when $H$ is constant one can transform the  McVittie metric into the static (and spherically symmetric) coordinate chart of the SdS spacetime.

The McVittie metric has been shown to provide a solution~\cite{AfshordiFontaniniGuariento2014} in certain subsets of Horndeski theory~\cite{Horndeski1974}, 
the most general scalar-tensor theory leading to 2nd order field equations in four dimensions.
It is thus natural to ask whether other SSS metrics exist which can be coordinately transformed into a 
McVittie type solution in some generic theory of gravity. 
To make the question more precise, let  $g[{\cal P}_i]$ be a metric with spherical symmetry, which depends on a set of parameters ${\cal P}_i$
such that when ${\cal P}_i \rightarrow 0$ the metric becomes FRW. What are the conditions under which  $g[{\cal P}_i]$  is equivalent to a
SSS metric?

It may be shown that a general FRW spacetime cannot be transformed into a static spacetime except in the case of pure de Sitter (dS)~\footnote{
For brevity, we will be calling ``de Sitter'' a general class of spacetimes with cosmological constant $\Lambda$ of any value, i.e. positive, negative or zero,
that is, including anti-de Sitter and Minkowski.  }. This is
easy to understand as a FRW spacetime has 6 Killing vectors associated with translations and rotations
while a SSS spacetime has 4  Killing vectors associated with rotations and time-translations, thus the two types of spacetime cannot be coordinate equivalent.
The exception to this rule is the dS spacetime which has 10 Killing vectors and which is known to be expressible in
both a SSS form, or a FRW form~\cite{Schrodinger1956}. See ~\cite{Ibison2007} for more explicit discussion.
Hence, the SSS metric we seek, $g_{\SSS}[{\cal P}_i]$,  must become pure dS when ${\cal P}_i \rightarrow 0$.

Unfortunately, the solution to this problem is very difficult, if not intractable, in general.
Fortunately, in the case where the  SSS spacetime is approximately dS, it can be transformed, as show in this article,
into an approximately FRW spacetime provided certain conditions are met. That is, let $g[{\cal P}_i] = g_{\FRW} + \delta g_{FRW}[{\cal P}_i]$
where $\delta g_{\FRW}[{\cal P}_i]$ is a small perturbation around FRW which vanishes when ${\cal P}_i \rightarrow 0$
and likewise let $g_{\SSS}[{\cal P}_i] = g_{\dS} + \delta g_{\SSS}[{\cal P}_i]$ where   $\delta g_{\SSS}[{\cal P}_i]$ 
 is a small perturbation around dS which vanishes when ${\cal P}_i \rightarrow 0$. 
Then $g[{\cal P}_i]$ is equivalent to $g_{\SSS}[{\cal P}_i]$ provided certain conditions that we determine in  this article are met.


This article is organised as follows. In section \ref{section_perturbed_FRW} we give a short overview of perturbed FRW spacetimes as relevant 
to this work and give particular emphasis to the existence of zero-modes which prohibit the separation of a metric perturbation 
into gauge-invariant and gauge-variant parts. In section \ref{transforming_SSS} we present the steps for determining the conditions for 
transforming a SSS metric which is approximately dS into an approximately FRW metric and vice versa. Our result is captured by 
\eqref{static_Phi_solution} and \eqref{static_Psi_solution}. In  section \ref{turnaround} we present a simple application of our construction,
 namely the calculation of the turnaround radius which is the distance away from a spherical mass distribution
 where the attractive force on a test particle due to the central mass  is balanced by the repulsion due to the presence of dark energy.
We conclude in section \ref{conclusion}. 

Throughout the article we use a $-+++$ metric signature convention and use units where the speed of light is unity. In addition, we use
the greek alphabet for spacetime indices and the latin alphabet for spatial indices.

\section{Perturbed FRW spacetimes}
\label{section_perturbed_FRW}
The general perturbation of the  FRW metric, involving only scalar modes, is 
\begin{equation}
 ds^2 = - (1 +2\psi) dt^2 -  2   \grad_i \zeta  \; dtdx^i 
+ a^2 \left[ \left( 1 +  \frac{1}{3} h \right) \gamma_{ij} + D_{ij} \nu \right] dx^i dx^j
\label{perturbed_FRW}
\end{equation}
where $D_{ij} = \grad_i \grad_j - \frac{1}{3} \grad^2 \gamma_{ij}$ is a  traceless derivative operator 
and  $\psi$, $\zeta$, $h$ and $\nu$ are the four possible scalar modes. We ignore vector and tensor modes as they are not relevant to this work.
The metric \eqref{perturbed_FRW} is subject to gauge transformations generated by the vector field $\xi^\mu =  (\xi_T, \grad^i \xi_L)$, where
$\grad^i = \gamma^{ij} \grad_j$ and where we have kept only the scalar modes which are part of the vector field $\xi^\mu$.
We choose to express the four scalar modes in terms of the Newtonian gauge perturbations $\Phi$ and $\Psi$
plus the two scalar gauge modes $\xi_T$ and $\xi_L$ as
\begin{align}
 \psi &= \Psi  + \dot{\xi}_T
\label{eq_psi_GI}
\\
\grad_i \zeta &=   \grad_i \left( \xi_T - a^2 \dot{\xi}_L \right)
\label{eq_zeta_GI}
\\
h  &=  -6\Phi + 6 H \xi_T + 2 \grad^2 \xi_L
\label{eq_h_GI}
\\
D_{ij}\nu &=   2  D_{ij}  \xi_L 
\label{eq_nu_GI}
\end{align}
In Fourier space the two potentials $\Phi$ and $\Psi$ are gauge-invariant
and it can be shown that it is always possible to split the metric perturbation into a gauge-invariant 
and a gauge-variant part~\cite{Nakamura2010,Nakamura2013}.  

Unfortunately, in real space the presence of zero-modes i.e. modes belonging to the kernel of the operators $\grad_i$ and $D_{ij}$,
 prohibits the complete separation of the metric perturbation in this way (see also ~\cite{Nakamura2012}). If we ignore these two operators, which 
amounts to ignoring the zero-modes, then we can eliminate the gauge transformation modes $\xi_T$ and $\xi_L$ from the above relations and
solve for the potentials $\Phi$ and $\Psi$ as
\begin{align}
\Phi  &= -\frac{1}{6} \left( h - \grad^2 \nu \right)  +  H \left( \zeta + \frac{1}{2} a^2 \dot{\nu} \right)
\label{Phi_def}
\\
 \Psi &= \psi  - \dot{\zeta} - \frac{1}{2} a^2 \left(\ddot{\nu} + 2 H \dot{\nu}\right)
\label{Psi_def}
\end{align}
The existence of zero-modes means that under the transformation 
\begin{align}
\zeta &\rightarrow \zeta +  \xi_0(t)
\\
\nu &\rightarrow \nu + \xi_1(t) + \frac{1}{2} \xi_2(t)  r^2,
\end{align}
where $\xi_0$, $\xi_1$ and $\xi_2$ are functions of $t$ only, $\Phi$ and $\Psi$ (and hence, the Newtonian gauge) are neither unique
nor gauge-invariant but they transform as 
\begin{align}
\Phi  &\rightarrow  \Phi +  H \xi_0 + \frac{1}{2}  \xi_2 +   \frac{1}{2} H a^2 \left(\dot{\xi}_1 + \frac{1}{2} \dot{\xi}_2  r^2\right) 
\label{Phi_kernel}
\\
\Psi &\rightarrow  \Psi -   \dot{\xi}_0 
 - \frac{1}{2}  a^2 \left(\ddot{\xi}_1  + 2   H \dot{\xi}_1   \right)
- \frac{1}{4} a^2  \left( \ddot{\xi}_2  +  2  H \dot{\xi}_2  \right) r^2 
\label{Psi_kernel}
\end{align}
Given any general $\Phi$ and $\Psi$, we can always construct
$\PhiGI$ and $\PsiGI$ with the zero modes removed as follows. Identify the purely time-dependent part of $\Phi$ as
$C(t) =  H \xi_0 + \frac{1}{2}  \xi_2 +   \frac{1}{2} H a^2 \dot{\xi}_1$ and the  purely time-dependent part of $\Psi$
as $E(t) = -   \dot{\xi}_0 - \frac{1}{2}  a^2 \left(\ddot{\xi}_1  + 2   H \dot{\xi}_1   \right)$.  Then
$\frac{1}{2H} \left(\dot{\xi}_2 - \frac{\dot{H}}{H} \xi_2\right) = \frac{\dot{C}}{H} - \frac{\dot{H} C }{H^2} + E$ which may be solved to get
$\xi_2 =  2 C + 2 H \int E dt$ and  where we have ignored any integration constants as they are irrelevant. 
Hence, $\xi_2$ is fully specified by the  purely time-dependent parts of $\Phi$ and $\Psi$ which in turn, fully specifies the parts
proportional to $r^2$. One may then proceed to subtract both of these parts from $\Phi$ and $\Psi$ in order to form $\PhiGI$ and $\PsiGI$.
We will call  $\PhiGI$ and $\PsiGI$ as the canonical form of the Newtonian gauge.
\label{kernel_section}

We shall return to this point further below, when we consider the conditions for the existence of a coordinate transformation between the 
SSS spacetime and the perturbed FRW spacetime.

\section{Transforming SSS metrics into perturbed FRW spacetimes}
\label{transforming_SSS}

\subsection{Transformation of a  general SSS metric}
Consider a  general SSS metric
\begin{equation}
 ds^2_{SSS} = -f(R) dT^2 + h(R) dR^2 + R^2 d\Omega
\label{gen_static_metric}
\end{equation}
where $f(R)$ and $h(R)$ are two functions of the radial coordinate $R$. We would like to determine the necessary conditions such that \eqref{gen_static_metric}
may be transformed into a perturbed FRW metric. 

Clearly, the first requirement is that the  perturbed FRW metric also has spherical symmetry, meaning that
all scalar potentials are functions of $t$ and $r$ only, i.e. $\Psi = \Psi(t,r)$
 and likewise for the other potentials. Then the metric \eqref{perturbed_FRW} may be casted as
\begin{align}
 ds^2 &= - (1 +2\Psi + 2\dot{\xi}_T ) dt^2 -  2 \frac{\partial}{\partial r} \left( \xi_T - a^2 \dot{\xi}_L   \right)  \; dtdr
\nonumber
\\
& 
+ a^2 \left( 1   - 2  \Phi + 2 H \xi_T + 2  \frac{\partial^2\xi_L}{\partial r^2} \right)  dr^2 
+ a^2 r^2 \left( 1 - 2 \Phi + 2 H \xi_T +  \frac{2}{r}  \frac{\partial\xi_L}{\partial r} \right)  d\Omega
\label{spher_perturbed_FRW}
\end{align}

If the two metrics \eqref{spher_perturbed_FRW} and \eqref{gen_static_metric} are to be related by a coordinate transformation, this 
suggests defining the radial coordinate $R$ via
\begin{equation}
R(t,r) =  \Rb \left[ 1 - \Phi +  H \xi_T +  \frac{1}{r} \frac{\partial\xi_L}{\partial r}  \right] 
\qquad
\qquad
\Rb \equiv a r
\end{equation}
while the other coordinate is casted as $T = T(t,r)$.
Taking derivatives and matching the two metrics we find
\begin{align}
 f \dot{T}^2 
=
  1 +2\Psi + 2\dot{\xi}_T 
 + h H^2 \Rb^2 \left[  1 + 2 \dot{\xi}_T   - \frac{2}{H} \dot{\Phi} +  2\frac{\dot{H} }{H}  \xi_T 
  +  \frac{2}{H r} \frac{\partial \dot{\xi}_L}{\partial r}    \right]
\label{dot_T_square},
\end{align}
\begin{align}
 \frac{1}{a^2}  f  \dot{T}  \frac{\partial T}{\partial r} 
& =
 h H r  
  \bigg[  1 -  2\Phi - \frac{1}{H} \dot{\Phi} - r \frac{\partial \Phi }{\partial r}
+   2H \xi_T 
+  \frac{\dot{H}}{H}  \xi_T + \dot{\xi}_T  
+  H r \frac{\partial \xi_T }{\partial r} 
\nonumber
\\
&
+  \frac{1}{Hr} \frac{\partial \dot{\xi}_L}{\partial r} 
+  \frac{1}{r} \frac{\partial\xi_L}{\partial r}  
+  \frac{\partial^2\xi_L}{\partial r^2} 
\bigg] 
 +  \frac{1}{a^2} \frac{\partial \xi_T }{\partial r} 
-  \frac{\partial \dot{\xi}_L}{\partial r} 
\label{dot_T_T_prime}
\end{align}
and
\begin{align}
 \frac{1}{a^2} f  \left( \frac{\partial T}{\partial r}  \right)^2 
= 
(h - 1 )  \left[ 1   - 2  \Phi + 2 H \xi_T + 2  \frac{\partial^2\xi_L}{\partial r^2} 
\right]  
+2  h  r \left[ H  \frac{\partial \xi_T }{\partial r} -   \frac{\partial \Phi }{\partial r} \right]
\label{T_prime_square}
\end{align}

\subsection{Transforming an  approximately de Sitter SSS metric}
Let us assume that we have a perturbed FRW spacetime as in \eqref{perturbed_FRW} but with the background being exact de Sitter, 
i.e. $a\rightarrow \abar = e^{\Hb t}$ with  $H  \rightarrow \Hb = \sqrt{\frac{\Lambda}{3}}$ a constant. Likewise,
let us assume that the static spacetime is approximately de Sitter and let 
\begin{align} 
f(R) &= \fb \left(1+ 2\lambda\right) = \left(1 - \Hb^2 R^2\right) \left(1 + 2 \tilde{\lambda}\right)
\\
h(R) &= \fb^{-1}\left(1 + 2\sigma\right) =  \frac{1 + 2 \tilde{\sigma}}{1 - \Hb^2 R^2} 
\end{align}
where $\fb = 1   - \Hb^2 \Rb^2$ and the perturbations $\lambda(\Rb)$,  $\tilde{\lambda}(\Rb)$, $\tilde{\sigma}(\Rb)$ 
and $\sigma(\Rb)$ are functions of $\Rb$ only (or equivalently of $R$ to this order). These definitions imply the relations
 $\fb \tilde{\sigma} =  \fb \sigma - \Hb^2 \Rb^2 \delta_R$ and $\fb\tilde{\lambda} =  \fb \lambda + \Hb^2 \Rb^2 \delta_R$, 
where $\delta_R$ is defined through $R = \bar{R} (1 + \delta_R)$. On the SSS spacetime $\delta_R$ can always be
set to zero via a radial coordinate transformation.

With these  assumtions \eqref{dot_T_square} and \eqref{T_prime_square} lead to~\footnote{Strictly
 speaking there is a sign occuring in each of \eqref{T_dot_alt} and \eqref{T_r_alt} when taking the square root.
Using \eqref{dot_T_T_prime} however, it turns out that both signs must be equal and by convention may be chosen to be positive.}
\begin{align}
 \dot{T}
=
\Psi +  \dot{\xi}_T
+ \frac{1}{\fb} \left[ 1 -  \tilde{\lambda} 
 +   \Hb^2 \Rb^2   \left(  \tilde{\sigma} 
+ \frac{2}{\fb} \delta_R 
+ \frac{1}{\Hb}  \dot{\delta}_R 
 \right)
\right]
\label{T_dot_alt}
\end{align}
and
\begin{align}
 \frac{\partial T}{\partial r}  
= 
 \frac{\abar \Hb \Rb }{\fb} \left( 1 
- \tilde{\lambda}
+ \frac{1}{\Hb^2 \Rb^2} \tilde{\sigma}
+ \frac{2}{\fb} \delta_R
+ \frac{r}{ \Hb^2 \Rb^2} \frac{\partial\delta_R }{\partial r}  
\right)  
+ \frac{1}{\Hb r}  \left( \frac{1}{r}  \frac{\partial\xi_L}{\partial r} - \frac{\partial^2\xi_L}{\partial r^2} \right)
\label{T_r_alt}
\end{align}
%
respectively. In addition, \eqref{dot_T_T_prime} leads to the consistency condition
\begin{align}
   \fb\tilde{\sigma}
&= 
 r\frac{\partial \Phi }{\partial r}
 - \Hb^2 \Rb^2\left[ \Psi 
 + \frac{1}{\Hb} \dot{\Phi} 
\right]
\label{cross_1}
\end{align}
which is independent of the gauge-fixing terms $\xi_T$ and $\xi_L$ and is also completely gauge-invariant under \eqref{Phi_kernel} and \eqref{Psi_kernel}. 
This means that in the above relation we can replace $\Phi$ and $\Psi$ with their canonical forms. This was to be expected as the left hand side
of \eqref{cross_1}, i.e. $\fb\tilde{\sigma}$, should have no knowledge of the gauge used for the perturbed FRW metric.


Taking the $r$-derivative of  \eqref{T_dot_alt} and equating to the $t$-derivative of $\eqref{T_r_alt}$ gives us a futher consistency condition
\begin{align}
  \frac{\partial\tilde{\lambda}}{\partial r} 
-\abar \Hb \Rb \dot{\tilde{\lambda}}
=
  \fb  \frac{\partial}{\partial r}\left(  \Psi + \frac{1}{\Hb}   \dot{\Phi} \right)
- \frac{\abar}{\Hb \Rb} \dot{\tilde{\sigma}}
 + \Hb^2 \Rb^2  \frac{\partial\tilde{\sigma} }{\partial r} 
\label{rt_equals_tr}
\end{align}
which is also completely gauge-invariant under \eqref{Phi_kernel} and \eqref{Psi_kernel}.

We have found two gauge-invariant conditions for the coordinate equivalence of the two metrics. However, our job is not yet done.
We need to make sure that the perturbed FRW metric also has the same four Killing vectors 
as the SSS metric, that is,  the action of the Killing vector $\partial_T$ on the perturbed FRW metric gives zero.
Inverting the transformation matrix defined by \eqref{T_dot_alt},  \eqref{T_r_alt} and $dR$ gives us the following relations
\begin{align}
\frac{\partial t}{\partial T} &= 1 -  \Psi 
+  \tilde{\lambda} 
+  \tilde{\sigma}
- \dot{\xi}_T
+  \frac{1}{r}  \frac{\partial\xi_L}{\partial r} - \frac{\partial^2\xi_L}{\partial r^2} + r \frac{\partial \delta_R }{\partial r}
,
\\
\frac{\partial t}{\partial R} &=
 - \frac{\Hb \Rb }{ \fb } \bigg[
1 -  \Psi +  \tilde{\sigma} - \dot{\xi}_T + \frac{1}{\Hb^2 \Rb^2} \tilde{\sigma}
+ \frac{2}{\fb} \delta_R
+ \frac{r}{ \Hb^2 \Rb^2} \frac{\partial \delta_R }{\partial r} 
\nonumber
\\
&
+ \frac{1}{ \Hb^2 \Rb^2 }  \left( \frac{1}{r}  \frac{\partial\xi_L}{\partial r} - \frac{\partial^2\xi_L}{\partial r^2} \right)
 \bigg]
,
\\
\frac{\partial r}{\partial T} &=
- \Hb r \left[ 1 -  \Psi +  \tilde{\lambda} +  \tilde{\sigma}
- \dot{\xi}_T +  \frac{1}{r}  \frac{\partial\xi_L}{\partial r} - \frac{\partial^2\xi_L}{\partial r^2} 
+ \frac{1}{\Hb}\dot{\delta}_R \right] 
,
\\
\frac{\partial r}{\partial R} &=
 \frac{1}{\abar \fb} \bigg[ 
 1 -  \Hb^2 \Rb^2 \Psi 
+  \tilde{\sigma}
- \Hb^2 \Rb^2 \dot{\xi}_T
+  \frac{1}{r}  \frac{\partial\xi_L}{\partial r} - \frac{\partial^2\xi_L}{\partial r^2}
\nonumber
\\
&
 +   \Hb^2 \Rb^2   \left(  \tilde{\sigma} + \frac{2}{\fb} \delta_R + \frac{1}{\Hb}  \dot{\delta}_R \right)
\bigg]
\end{align}
Using the above relations, we transform the Killing vector: $K^\mu = \partial_T$ into the FRW system. Its  components are
\begin{align}
K^t &= 1 -  \Psi +  \tilde{\lambda} +  \tilde{\sigma} - r \frac{\partial\Phi}{\partial r} - \dot{\xi}_T + \Hb r \frac{\partial \xi_T}{\partial r} 
,
\\
K^r &= - \Hb r \left( 1 -  \Psi +  \tilde{\lambda} +  \tilde{\sigma}
- \dot{\xi}_T +  \frac{1}{r}  \frac{\partial\xi_L}{\partial r} - \frac{\partial^2\xi_L}{\partial r^2} 
+ \frac{1}{\Hb}\dot{\delta}_R \right)
\end{align}
Acting with the Killing vector field $K$ on $g_{\mu\nu}$ in the FRW coordinate system gives the following three conditions
\begin{align}
  r \frac{\partial}{\partial r}  \left( \Hb \Psi + \dot{\Phi} \right)
-  \frac{\partial}{\partial t} \left( \tilde{\lambda} +  \tilde{\sigma}  \right)
 &= 0 
,
\label{kill_1}
\\
    \frac{\partial}{\partial r} \left[   \Psi +r \frac{\partial\Phi}{\partial r}-  \tilde{\lambda} -  \tilde{\sigma} \right]
+  \abar^2 \Hb r   \frac{\partial}{\partial t} \left[   \Psi  + \frac{1}{\Hb} \dot{\Phi}  -  \tilde{\lambda} -  \tilde{\sigma}\right]
&=0
,
\label{kill_2}
\\
 \frac{\partial }{\partial r} \left(   \Psi  + \frac{1}{\Hb} \dot{\Phi} - \tilde{\lambda} -  \tilde{\sigma} \right)
&=0
\label{kill_3}
\end{align}

Equation \eqref{kill_3} may be integrated to give
\begin{equation}
 \Psi  + \frac{1}{\Hb} \dot{\Phi} - \tilde{\lambda} -  \tilde{\sigma} = \alpha
\label{int_kill}
\end{equation}
where $\alpha = \alpha(t)$. 
Combining \eqref{kill_1} and \eqref{kill_2} we get
\begin{align}
 \Hb r \frac{\partial}{\partial r}  \left(  \tilde{\lambda} +  \tilde{\sigma} \right)
 &= \frac{\partial}{\partial t} \left( \tilde{\lambda} +  \tilde{\sigma}  \right)
\end{align}
which  is equivalent to saying that $\tilde{\lambda} +  \tilde{\sigma} = {\cal A}(\Rb) $ is a function of $\Rb$ only. 
Incidentally, this means that
\begin{align}
 \Psi  + \frac{1}{\Hb} \dot{\Phi} = {\cal A}(\Rb) + \alpha
\label{int_kill_2}
\end{align}
Using \eqref{int_kill} into \eqref{kill_2} and integrating gives
\begin{align}
-  \dot{\Phi}
+ \Hb r \frac{\partial\Phi}{\partial r} &= \Hb \dot{\beta} - \frac{1}{2} \Hb^2 \Rb^2 \dot{\alpha}
\end{align}
where $\beta =\beta(t)$ only.
This implies that 
\begin{align}
\Phi = \Phi_0  - \Hb\beta +    \frac{1}{2}  \alpha\Hb^2  \Rb^2+ {\cal B}(\Rb)
\label{static_Phi_solution}
\end{align}
where  $\Phi_0$ is a  constant and ${\cal B}(\Rb)$ is an arbitrary function of $\Rb$ which is determined by the field
equations of the theory in question.
Finally, using  \eqref{static_Phi_solution} into \eqref{int_kill_2} we find an equivalent condition for $\Psi$ which is
\begin{align}
 \Psi  = \dot{\beta}  + \alpha
-  \left(   \alpha\Hb^2   + \frac{1}{2}  \dot{\alpha} \Hb \right) \Rb^2
 + {\cal A}(\Rb) 
- \Rb   \frac{d{\cal B}}{d\Rb} 
\label{static_Psi_solution}
\end{align}
We have shown that in order for a perturbed FRW metric to be coordinate equivalent to a SSS metric,
the Newtonian gauge potentials must have the form \eqref{static_Phi_solution} and \eqref{static_Psi_solution} respectively.

As we have already discussed, the Newtonian potentials $\Phi$ and $\Psi$ are not gauge-invariant. Notice, however, that
the obtained functional forms \eqref{static_Phi_solution} and \eqref{static_Psi_solution} have precicely the same structure as the 
gauge-transformations of the potentials in \eqref{Phi_kernel} and \eqref{Psi_kernel}. 
Indeed, following the procedure outlined in section \ref{kernel_section}, we may recast the potentials into canonical form. In particular
the constant $\Phi_0$ and the two functions $\alpha(t)$ and $\beta(t)$ are gauge artifacts and the canonical forms of the potentials are
\begin{align}
\PhiGI = {\cal B}(\Rb)
\label{Phi_GI_solution}
\end{align}
and
\begin{align}
 \PsiGI  = {\cal A}(\Rb) - \Rb   \frac{d{\cal B}}{d\Rb} 
\label{Psi_GI_solution}
\end{align}
respectively. The functions ${\cal A}(\Rb)$ and ${\cal B}(\Rb)$ are determined by the field equations of the gravitational theory that we 
may apply this procedure to.
Choosing the Newtonian gauge and furthermore bringing the potentials into canonical form
 fully determines the functions $\dot{T}$ and $\frac{\partial T}{\partial r}$ in \eqref{T_dot_alt} and \eqref{T_r_alt}, which in turn
determine the coordinate transformation completely.

We now describe how to transform a given metric satisfying the above conditions from the FRW into the SSS system
and vice-versa.

\subsubsection{From the cosmological to the static space}
\label{cosmo_static}
Given $\Phi$ and $\Psi$ in the form \eqref{static_Phi_solution} and \eqref{static_Psi_solution} we  use
 \eqref{cross_1} to obtain $\tilde{\sigma}$.  As this expression is  gauge-invariant, we may use it in any gauge of choice, not necessarily 
the Newtonian gauge, by substituting $\Phi$ and $\Psi$ with the expressions \eqref{Phi_def} and \eqref{Psi_def}. Alternatively we may substitute 
$\Phi$ and $\Psi$ with their canonical expressions and use \eqref{Phi_GI_solution} and \eqref{Psi_GI_solution} to get
\begin{align}
\tilde{\sigma}  &= \frac{1}{\fb}\left[ r\frac{\partial \Phi }{\partial r} - \Hb^2 \Rb^2\left( \Psi + \frac{1}{\Hb} \dot{\Phi} \right) \right]
  \rightarrow   \frac{1}{\fb}\left[  \Rb \frac{d{\cal B}}{d\Rb} - \Hb^2 \Rb^2{\cal A}   \right]
\label{sigma_tilde_of_Phi_Psi}
\end{align}

In order to get $\tilde{\lambda}$ we use $\tilde{\lambda} = {\cal A} - \tilde{\sigma}$. The function $ {\cal A}$ may be obtained 
after transforming the potentials into canonical form, or, by considering the combination
$\Psi + \frac{1}{\Hb} \dot{\Phi}$ and subtracting the purely time-dependent function $\alpha(t)$. 
This leads to
\begin{align}
\tilde{\lambda} &\rightarrow
\frac{1}{\fb}\left[  {\cal A}  - \Rb \frac{d{\cal B}}{d\Rb}  \right]
=  \frac{1}{\fb} \PsiGI
\label{lambda_tilde_of_Phi_Psi}
\end{align}
It is easy to check that the above relations satisfy \eqref{rt_equals_tr}.

\subsubsection{From the static to the cosmological space}
\label{static_cosmo}
Now consider the inverse transformation. Suppose we have at hand $\tilde{\sigma}$ and $\tilde{\lambda}$ as a function of $\Rb$ and we want to 
determine the cosmological metric. 
Of course, in order to do so, we need to choose in which gauge we want to perform the mapping into.
It is simpler if we first determine $\Phi$ and $\Psi$, and then adopt them in the gauge of choice. However,
 even $\Phi$  and $\Psi$ are not invariant, hence, we firstly, determine the canonical forms $\PhiGI$ and $\PsiGI$
 by inverting \eqref{sigma_tilde_of_Phi_Psi} and \eqref{lambda_tilde_of_Phi_Psi}
to get
\begin{align}
\PhiGI &=  \int^{\Rb} \frac{d\Rb}{\Rb} \left[ \tilde{\sigma}+ \Hb^2 \Rb^2 \tilde{\lambda}\right]
\label{B_of_sigma_lambda}
\end{align}
and
\begin{align}
 \PsiGI  &=  \fb  \tilde{\lambda} 
\label{A_of_sigma_lambda}
\end{align}
We may then perform gauge-transformations according to \eqref{Phi_kernel} and \eqref{Psi_kernel} followed by a specific gauge choice.

\subsubsection{FRW backgrounds  close to de Sitter}
\label{close_de_sitter}
Suppose that we are given a perturbed FRW spacetime with scale factor $a$ and solutions in the Newtonian gauge given by $\tilde{\Phi}$ and $\tilde{\Psi}$.
Suppose further that the background FRW is close to de Sitter so that $a = \abar(1 + \delta_a(t))$. What are the conditions on
$\tilde{\Phi}$ and $\tilde{\Psi}$ in order for this spacetime to be coordinate equivalent to a SSS spacetime?
Given $\tilde{\Phi}$ and $\tilde{\Psi}$, it is always possible to determine their canonical form $\hat{\tilde{\Phi}}$ and $\hat{\tilde{\Psi}}$ via
 the procedure in \ref{kernel_section}. It is therefore suffiecient to determine the conditions on the canonical forms.

If the background FRW is close to de Sitter then we may define a potential $\Phi = \tilde{\Phi} - \delta_a$ 
 such that $a^2(1 - 2\tilde{\Phi})\gamma_{ij} = \abar^2(1 - 2\Phi)\gamma_{ij}$. In this way, our
 perturbed FRW spacetime with background approximately de Sitter is equivalent to a perturbed de Sitter spacetime with
potentials $\Phi$ and  $\Psi = \tilde{\Psi}$. We have already determined the conditions for such a spacetime to be coordinate equivalent to a
SSS spacetime; they are the conditions given by \eqref{static_Phi_solution} and \eqref{static_Psi_solution}. It is 
  also possible to transform those conditions into canonical form, i.e. two 
are arbitrary functions $\PhiGI(\Rb)$ and $\PsiGI(\Rb)$ of $\Rb$. Hence we may write
\begin{align}
\tilde{\Phi} &= \delta_a(t) + \PhiGI(\Rb)
\\
\tilde{\Psi} &=  \PsiGI(\Rb)
\end{align}
We now determine, the canonical form of $\tilde{\Phi}$ and $\tilde{\Psi}$. Using the procedure outlined in \ref{kernel_section} by setting
 $C = \delta_a $ and $E=0$ we find $\xi_2= 2 \delta_a$. Hence, adding and subtracting 
appropriate terms  we find
\begin{align}
\hat{\tilde{\Phi}} &=  \PhiGI(\Rb) - \frac{1}{2} \Hb\dot{\delta}_a   \Rb^2 
\label{tilde_Phi_trans}
\\
\tilde{\Psi} &= \PsiGI(\Rb) + \frac{1}{2}  \left(\ddot{\delta}_a  +  2  \Hb \dot{\delta}_a  \right) \Rb^2 
\label{tilde_Psi_trans}
\end{align}

\section{An application: the turnaround radius}
\label{turnaround}
An interesting application of the construction considered in this article concerns the turnaround radius, 
the scale where the attraction due to the mass of a bound structure is balanced by the repulsion due to a component of dark energy.
The turnaround radius was calculated in~\cite{Stuchlik1983,StuchlikHledik1999} in the wider context 
of geodesics of the SdS spacetime and in~\cite{PavlidouTomaras2013} and~\cite{PavlidouTetradisTomaras2014} 
in the cosmological $\Lambda$CDM and smooth dark energy context.
In~\cite{BhattacharyaDialektopoulosTomaras2015} the turnaround radius  was calculated in the cubic galileon gravitational theory
while the turnaround radius in generic theories of gravity was tackled in~\cite{Faraoni2015} and~\cite{BhattacharyaEtAl2016}.

The turnaround radius is most easily defined in the case where the metric is SSS in which case it is given by the value of $R$  where $\frac{df}{dR}$
vanishes.
Substituting the case where $f$ is a perturbation on de Sitter we find
\begin{equation}
  (1 - \Hb^2 R^2) \frac{d \tilde{\lambda}  }{dR}  = \Hb^2 R (1 + 2 \tilde{\lambda} ) 
\label{turnaround_SSS}
\end{equation}
Thus when the function $\tilde{\lambda}$ is known (for instance after solving the field equations of a certain gravitational theory), the 
turnaround radius can be readily calculated by solving the above equation for $R$.

Consider, now the same problem as viewed from cosmology. Using $\tilde{\lambda} = \Psi/\fb$, \eqref{turnaround_SSS}  turns into
\begin{equation}
 \frac{d\Psi}{dR} = \Hb^2 R,
\end{equation}
however, changing to the cosmological coordinates $t$ and $r$ as well as using the fact that $d\Psi/dT=0$  we find
\begin{equation}
\frac{\partial \Psi}{\partial r} = \abar \Hb^2 R 
\end{equation}
When the function $\Psi(t,r)= \Psi(R)$ is known, the turnaround radius can be  readily calculated by solving the above equation for $R$.

The procudure above is valid only when the background FRW spacetime is exactly de Sitter. Based on our discussion in section \ref{close_de_sitter}
we may find the turnaround equation in the case when the background FRW spacetime is approximately de Sitter. Following the procedure given 
in  section \ref{close_de_sitter} we find  
\begin{equation}
r \frac{\partial \tilde{\Psi}}{\partial r} =  \left[H^2 + \dot{H}  \right] R^2
\label{turnaround_cosmo}
\end{equation}
The turnaround equation in~\cite{BhattacharyaEtAl2016} was derived 
in a different way and in a more general setting where spherical symmetry is not assumed initially. 
Our turnaround equation \eqref{turnaround_cosmo} agrees with~\cite{BhattacharyaEtAl2016} in the case of spherical symmetry.

\section{Conclusion}
\label{conclusion}

In this article we have determined the conditions under which a static spherically symmetric metric is equivalent to an approximately FRW metric.
Our result is captured by \eqref{static_Phi_solution} and \eqref{static_Psi_solution} which give the general form that the FRW metric potentials
can have for the equivalence to hold. We gave a prescription for transforming spherically symmetric perturbations of FRW 
to spherically symmetric perturbations de Sitter in section \ref{cosmo_static} and the opposite in  \ref{static_cosmo}
and considered what happens when the background cosmology is approximately de Sitter in \ref{close_de_sitter}.
Finally, we applied our construction to the simple example of the calculation of the turnaround radius. 

We close the article by a  conjecture generalizing the interpretation of the exact McVittie in GR as 
a black hole in an expanding universe when it asymptotes to de Sitter, as  was shown  by Kaloper,Kleban and Martin~\cite{KaloperKlebanMartin2010}.
That is, in order for a McVittie type solution of a generic theory of gravity 
to describe a black hole in an expanding universe it must be expressible in a perturbed FRW form with the
FRW background to be approximately de Sitter (and tend asymptotically to it) and the Newtonian gauge perturbations be expressible by
 \eqref{static_Phi_solution} and \eqref{static_Psi_solution} (within the considerations of section \ref{close_de_sitter}). Of course, further conditions whould most likely  be necessary, such as, the existence
of a black hole horizon when expressed in static coordinates.

\section*{Acknowledgements}
The research leading to these results has received funding from the European Regional Development Fund 
and the Czech Ministry of Education, Youth and Sports (MSMT) (Project CoGraDS - CZ.02.1.01/0.0/0.0/15\_003/0000437). 

\bibliographystyle{eprint_unsrt}
\bibliography{references}

\end{document}